\documentstyle[epsfig,aps,prl,twocolumn]{revtex}
\begin{document}

\wideabs{

\title{Direct evidence for a multiple superconducting gap in MgB$_{2}$ from high-resolution photoemission spectroscopy}

\author{S. Tsuda$^{1}$, T. Yokoya$^{1}$, T. Kiss$^{1}$, Y.Takano$^{2}$, K. Togano$^{2}$, H. Kitou$^{3}$, H. Ihara$^{3}$, and S. Shin$^{1,4}$}

\address{$^{1}$ Institute for Solid  State  Physics, University of Tokyo, Kashiwa, Chiba 277-8581, Japan}
\address{$^{2}$ National Institute for Materials Science, 1-2-1 Sengen, Tsukuba 305-0047, Japan}
\address{$^{3}$ National Institute of Advanced Industrial Science and Technology, 1-1-4 Umezono, Tsukuba 305-8568, Japan}
\address{$^{4}$ The Institute of Physical and Chemical Research (RIKEN), Sayo-gun, Hyogo 679-5143, Japan}

\date{\today}

\maketitle

\begin{abstract}\\
\hspace{0.5cm}We study the new binary intermetallic superconductor MgB$_{2}$ using high-resolution photo-emission spectroscopy. The superconducting-state spectrum measured at 5.4 K shows a coherent peak with a shoulder structure, in sharp contrast to that expected from a simple isotropic-gap opening. The spectrum can be well reproduced using the weighted sum of two Dynes functions with the gap sizes of 1.7 and 5.6 meV. Temperature-dependent study shows that both gaps close at the bulk transition temperature. These results provide spectroscopic evidence for a multiple gap of MgB$_{2}$.
\end{abstract}
}

MgB$_{2}$ has attracted tremendous attention after the discovery of superconductivity by Akimitsu {\it et al.}, because the simple intermetallic compound shows a remarkably high transition temperature ({\it T$_{c}$}) of 39 K (Ref. \cite{1}).  The value is actually the highest within intermetallic compounds and it is even higher than those of some cuprate high temperature superconductors, where paring driving forces other than phonon have been speculated \cite{2}.  For MgB$_{2}$, symmetry of the Cooper pairs has been confirmed to be of a spin-singlet type from nuclear magnetic resonance (NMR) studies \cite{3,4}. The boron isotope effect measurements \cite{5} indicate phonon-mediated superconductivity.  The next question to be address is why MgB$_{2}$ shows such a high {\it T$_{c}$}; {\it T$_{c}$} lies on or beyond the estimated upper-limit of {\it T$_{c}$} for phonon-mediated superconductivity \cite{6}.

Band structure calculations [7-10] predict coexistence of 2-dimensional covalent in-plane (hole like B-2p$_{xy}$ or $\sigma$-band) and 3-dimensional metallic-type interlayer (electron like B-2p$_{z}$ band or $\pi$-band) conducting bands at the Fermi level ({\it E$_{F}$}) for a peculiar feature of MgB$_{2}$.  For explaining its large {\it T$_{c}$} within the BCS framework [8-10], a strong coupling of high frequency phonons originating in the in-plane boron vibrations with the density of states at {\it E$_{F}$} is a plausible scenario.  However, the value of {\it T$_{c}$} can not be reproduced, if one uses obtained electron-phonon coupling constant $\lambda$ (= 0.65 - 1) and the commonly accepted values for Coulomb pseudopotential $\mu^{*}$ (Refs. [8-11]), suggesting need for a theory beyond ordinary BCS.  On the other hand, alternative models [12-15] which propose pairing mechanism other than phonon have also appeared.  

Since the symmetry and magnitude of a superconducting gap reflect the nature of superconductivity, many experimental studies on the superconducting gap of MgB$_{2}$ have been performed.  Tunneling [16-18] and photoemission \cite{18} measurements, which directly measure superconducting electronic structures and hence the size of the gap, have reported the size of the superconducting gap ($\Delta$) of 2 - 5.9 meV (the reduced gap value 2$\Delta$/{\it k$_{B}$T$_{c}$} = 1.2 - 3.5) assuming an isotropic gap.  Optical measurements roughly estimated 2$\Delta$/{\it k$_{B}$T$_{c}$} = 2.6 from the energy position where normalized reflectivity changes \cite{19}.  For the symmetry of the gap, while some studies have reported an isotropic s-wave gap [3,4,21-23], evidence for a deviation from an expectation of a simple isotropic s-wave gap has been increasing [24-28].  Though the symmetry and size of the gap seems not to be settled down completely even within the same experimental probe, these conflicts might hint on the shape of the superconducting gap.  Indeed, the results from specific heat \cite{25} and microwave surface resistance \cite{27} measurements have been discussed in the light of a multiple gap.

In this letter, we report high-resolution photoemission results on high density MgB$_{2}$ samples.  The spectrum measured at 5.4 K reveals a characteristic spectral shape which can not be explained by a simple isotropic gap opening.  From fittings to the temperature dependence of the gaps, we find the spectral shape is well described with two Dynes functions with different gap values and both of the gaps close at the bulk transition temperature.  This gives rise to a reduced-gap value of 1.08 for smaller gap and that of 3.56 for larger one.  These results provide direct evidence that the superconducting gap of MgB$_{2}$ is not a simple isotropic one, but is rather a multiple gap.

High-density MgB$_{2}$ samples were prepared with a high-pressure synthesis, details of which are described elsewhere \cite{28}.  Magnetization measurements show a superconducting transition with an onset and a midpoint at 38.0 and 36.5 K, respectively.  Photoemission measurements were performed on a spectrometer built using a GAMMADATA-SCIENTA SES2002 electron analyzer and high-flux discharging lamp with a toroidal grating monochromator.  The total energy resolution (analyzer and light) used the He I$\alpha$ (21.212eV) resonance line was set to 3.8 meV. The sample temperatures were measured using a silicon-diode sensor mounted just close to it.  The base pressure of the spectrometer was better than 5 x 10$^{-11}$ Torr.  Samples were fractured {\it in-situ} to obtain clean surfaces and no spectral changes were observed within the measurements. Fermi energy of samples was referenced to that of a gold film evaporated onto the sample substrate and its accuracy is estimated to be better than 0.2 meV.

Figure 1 shows photoemission spectra of MgB$_{2}$ measured at 5.4 K (superconducting state, opened circles connected with a line) and 45 K (normal state, open squares connected with a line) with He I$\alpha$ resonance line.  In contrast to the normal-state spectrum having a Fermi-edge like structure, the superconducting-state spectrum shows a clear peak around 7 meV.  To see the superconducting-state spectrum more in detail, we expanded the same spectrum at 5.4 K near {\it E$_{F}$}, as shown in the inset of Fig. 1.  We observe an intensity maximum at $\sim$ 7 meV and a shift of the leading edge, indicative of the opening of a superconducting gap.  More importantly, we find a shoulder structure at 3.5 meV as indicated with an arrow.  These structures were not clearly observed in the recent photoemission study \cite{18} most probably due to difference in experimental procedures (energy resolution, measured temperature, procedure of getting clean surface) and/or quality of the sample.  The observed anomalous leading edge spectral shape is in sharp contrast to those of elemental metals, Pb and Nb, where resolution-limited leading edges are observed \cite{29}.  These results suggest a deviation from a simple isotropic s-wave gap.

In order to get further insight into the shape of the superconducting gap, we analyzed the experimental spectrum with the Dynes function \cite{30} multiplied by the Fermi-Dirac function of 5.4 K and convolved with a Gaussian having a full width at half maximum (FWHM) of the known instrumental resolution.  The Dynes function is a modified BCS function in the form of D(E, $\Delta$, $\Gamma$) = Re \{(E - i$\Gamma$)/[(E - i$\Gamma$)$^{2}$-$\Delta^{2}$]$^{1/2}$\}, where E is the energy and $\Gamma$ is the thermal broadening parameter \cite{30}.  Figure 2(a) is an example of fitting results with $\Delta$ = 3.4 meV and $\Gamma$ = 1.5 meV (solid line) together with the experimental data (open circles), where one can see the leading edge part is not reproduced well.  We actually found that both of the peak and shoulder structures were not fit at the same time using any sets of parameters.  Supposing that the shoulder structure comes from another gap, we try to fit using the weighted sum (D$_{L+S}$) of two Dynes functions for a larger gap (D$_{L}$) and a smaller one (D$_{S}$), D$_{L+S}$ = (1/(1+R))D$_{L}$(E, $\Delta_{L}$, $\Gamma$)+ (R/(1+R))D$_{S}$(E, $\Delta_{S}$, $\Gamma$), where R is an amplitude ratio of the smaller gap to the larger one.  Here, we used the same value of $\Gamma$ for the two Dynes functions for simplicity.  In Fig. 2(b), we show a result of fittings (D$_{L+S}$, solid line), together with Dynes functions for the larger gap ($\Delta_{L}$ = 5.6 meV, broken line) and the smaller gap ($\Delta_{S}$ = 1.7 meV, dotted line) with $\Gamma$ = 0.10 meV and R = 5.2.  It is evident that D$_{L+S}$ reproduces the experimental result for both of the peak and shoulder structures significantly.  We note the magnitude ratio of the two gaps at 5.4 K, $\Delta_{L}$/$\Delta_{S}$, is $\sim$ 3.3. These analyses show that the superconducting gap of MgB$_{2}$ is not a simple isotropic one, but rather contains two dominant components.  This is cosistent to the transport [24-28] and most recent optical \cite{31} measurements, which indicate the inconsistency with an ordinary s-wave gap and some of which suggest a multicomponent gap.

To see how the two gaps behave as a function of temperature, we analyzed temperature-dependent spectra with D$_{L+S}$ assuming that R is temperature independent.  Figure 3 shows obtained temperature-dependent gap values, where open and filled circles show the sizes of the larger and smaller gaps, respectively.   Theoretical temperature dependence of gaps with $\Delta$(0) = 1.7 and 5.6 meV are also shown with broken and dotted lines, respectively \cite{32}.  We find, while the temperature dependence of the smaller gap follows the BCS prediction, that of the larger gap decreases faster than the prediction.  We do not know the reason for the deviation so far, but the result is similar to that obtained from MgB$_{2}$/Ag and MgB$_{2}$/In junctions \cite{33}.  More importantly, both of the two gaps close at nearly the midpoint of {\it T$_{c}$} (36.5 K) obtained from the magnetization measurements.  From these results, we obtain the reduced gap size 2$\Delta$(5.4 K)/{\it k$_{B}$}{\it T$_{c}$} of 3.56 for the larger gap, which is nearly consistent to the mean field value of 3.52, and that of 1.08 for the smaller gap.  The value of the smaller gap agrees well with that from the tunneling measurements by Rubio-Bollinger \cite{15}, while the values from the other tunneling \cite{16,17}, photoemission \cite{18}, and optical measurements \cite{20} lie between the obtained values of smaller and lager gaps.

Currently, there are active discussions on the mechanism of the superconductivity of MgB$_{2}$, as briefly discussed above.  Within the weak or strong coupling BCS models, it seems necessary to think of a multiple gap \cite{34} in order to explain the high value of {\it T$_{c}$} using reasonable $\lambda$ and $\mu^{*}$ as well as experimentally observed physical properties [24-28].  Liu {\it et al.} \cite{34} calculated a temperature dependence of two gaps in the weak-coupling multiple-gap model, showing that the larger and smaller gaps close at the same temperature and the magnitude ratio of larger one to smaller one is approximately three.  These predictions can describe the present results very well.  Present results are also consistent to the alternative models proposing exisotic mechanism to explain large {\it T$_{c}$} (Refs. \cite{12,13}), as long as they predict a multicomponent gap.  Further detailed studies to show calculated density of states are desired so as to be compared with the present study.  We do not attempt to assign whether the experimentally observed larger and smaller gaps relate to the bands with what kind of character, $\sigma$ band or $\pi$ band, since there is a conflict on the assignment between models.  Most recently, using a weak-coupling anisotropic s-wave model, Haas and Maki \cite{35} determined the shape of the quasiparticle density of states having a cusp at a maximum gap value and an onset of spectral weight at the minimum gap value, reminiscent of the present spectrum at 5.4 K having the peak and shoulder structures.  Chen {\it et al.} \cite{36} accounted for their tunneling spectra using an anisotropic s-wave model.  It is hard, for the present study alone, to distinguish a multiple gap originating in the Fermi surfaces with different character from a gap anisotropy within each Fermi surface. To directly study the multicomponent gap and/or the anisotropy in relation to the bands, angle-resolved photoemission using single crystals are necessary and urgent.

In conclusion, we report on high-resolution photoemission results on the superconducting gap of a new binary intermetallic superconductor, MgB$_{2}$.  The superconducting-state spectrum measured at 5.4 K shows a coherent peak with a shoulder structure, in sharp contrast to that expected from a simple isotropic-gap opening.  We find that a simulation using two Dynes functions with the gap sizes of 1.7 and 5.6 meV reproduces the superconducting-state spectrum better than that using a single Dynes function.  We also find that the smaller gap as well as the larger gap closes at the bulk transition temperature.  These results are consistent to transport measurements, and thus indicate the superconducting gap of MgB$_{2}$ is a multiple gap. 

We thank Professors M. Imada and K. Yamaji for useful discussions.  This work was supported by grants from the Ministry of Education, Culture and Science of Japan.


\newpage

\begin{figure}
\centerline{\epsfig{file=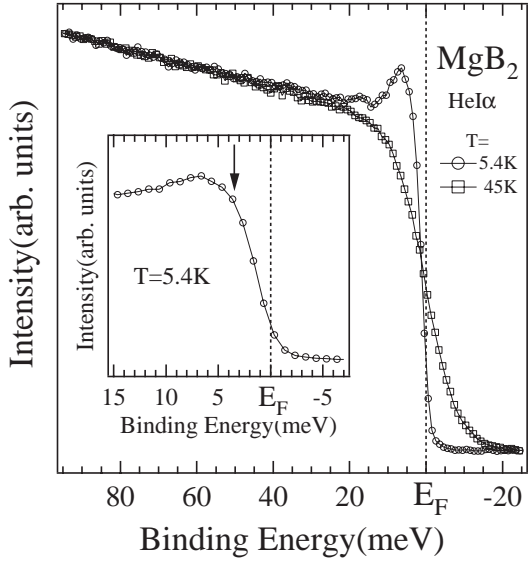,width=7cm}}
\hspace{0.5cm}
\caption{High-resolution photoemission spectra of MgB$_{2}$ measured at 5.4K (open circles connected with solid line) and 45K (open squares connected with solid line) with He I$\alpha$ resonance line (21.212eV).  The inset shows a expanded spectrum at 5.4K in the vicinity of {\it E$_{F}$}.  Please note that the spectrum has a peak with a shoulder structure as is emphasized with an arrow, which indicates non simple isotropic gap.}
\label{fig1}
\end{figure}

\begin{figure}
\centerline{\epsfig{file=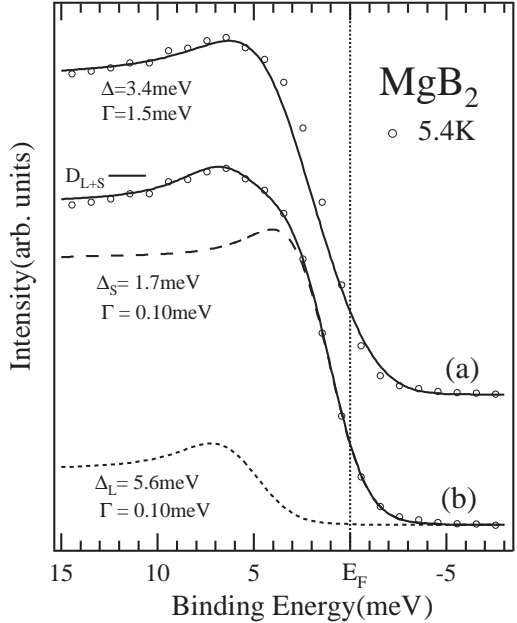,width=6.8cm}}
\hspace{0.5cm}
\caption{Results of fitting using (a) one Dynes function with $\Delta$ = 3.4 meV and $\Gamma$ = 1.5 meV (solid line) and (b) the weighted sum of two Dynes functions (solid line) with $\Delta_{S}$ = 1.7 meV (broken line) and $\Delta_{L}$ = 5.6 meV (dotted line) having the same $\Gamma$ = 0.10 meV.  Experimental spectrum at 5.4 K is shown with open circles.  For detailed fitting procedure, see text.}
\label{fig2}
\end{figure}

\begin{figure}
\centerline{\epsfig{file=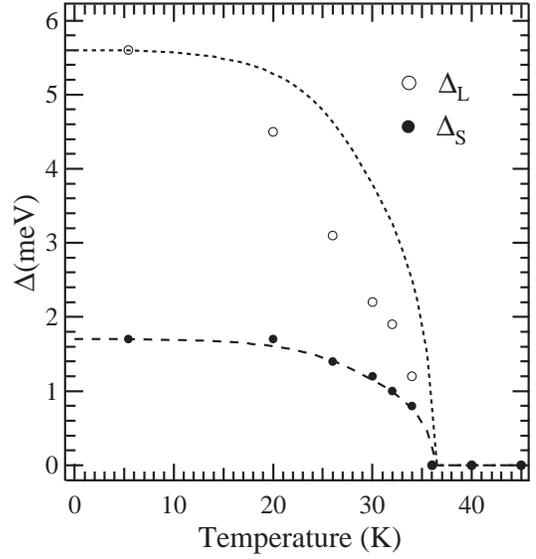,width=7cm}}
\hspace{0.5cm}
\caption{Temperature dependence of the two gaps obtained from the Dynes function analyses as described in the text.  Filled and open circles represent the sizes of the smaller and larger gaps, respectively.  Broken and dotted lines show the predicted temperature dependence of superconducting gap from BCS theory [33] for $\Delta$(0) = 1.7 and 5.6 meV, respectively.
}
\label{fig3}
\end{figure}

\end{document}